\documentclass[12pt]{article}
\usepackage[utf8]{inputenc}
\usepackage{amsmath,amssymb,bbm}
\usepackage{authblk}
\usepackage{cprotect}

\usepackage{stmaryrd}
\usepackage{colortbl,xcolor}

\usepackage{subcaption}

\usepackage{natbib}
\usepackage{booktabs, threeparttable}

\usepackage{geometry}
\usepackage{lipsum}
\usepackage{url} 
\usepackage{setspace}
\usepackage[figuresright]{rotating}
\usepackage{ dsfont }

\newtheorem{theorem}{Theorem}

\title{Nonparametric Bayesian approach for \\ dynamic borrowing of historical control data}
\author[1]{Tomohiro Ohigashi}
\author[2]{Kazushi Maruo}
\author[1]{Takashi Sozu}
\author[2]{Masahiko Gosho}
\affil[1]{Department of Information and Computer Technology, Faculty of Engineering, Tokyo University of Science, Tokyo, Japan}
\affil[2]{Department of Biostatistics, Institute of Medicine, University of Tsukuba, Tsukuba, Japan}
\date{\empty}
\date{\empty}

\begin{document}
\begin{singlespace}
\maketitle
% Corresponding author
\begin{flushleft}
	\textbf{Corresponding Author:}\\
	Tomohiro Ohigashi \\
	Department of Information and Computer Technology, Faculty of Engineering, Tokyo University of Science \\
	6-3-1 Niijuku, Katsushika-ku, Tokyo 125-8585, Japan\\
	Email: ohigashi@rs.tus.ac.jp
\end{flushleft}
\end{singlespace}

\begin{abstract}
When incorporating historical control data into the analysis of current randomized controlled trial data, it is critical to account for differences between the datasets.
When the cause of the difference is an unmeasured factor and adjustment for observed covariates only is insufficient, it is desirable to use a dynamic borrowing method that reduces the impact of heterogeneous historical controls.
We propose a nonparametric Bayesian approach for borrowing historical controls that are homogeneous with the current control.
Additionally, to emphasize the resolution of conflicts between the historical controls and current control, we introduce a method based on the dependent Dirichlet process mixture. 
The proposed methods can be implemented using the same procedure, regardless of whether the outcome data comprise aggregated study-level data or individual participant data.
We also develop a novel index of similarity between the historical and current control data, based on the posterior distribution of the parameter of interest.
We conduct a simulation study and analyze clinical trial examples to evaluate the performance of the proposed methods compared to existing methods.
The proposed method based on the dependent Dirichlet process mixture can more accurately borrow from homogeneous historical controls while reducing the impact of heterogeneous historical controls compared to the typical Dirichlet process mixture. 
The proposed methods outperform existing methods in scenarios with heterogeneous historical controls, in which the meta-analytic approach is ineffective.
\end{abstract}


\section{Introduction}\label{intro}
Clinical trials increasingly use information from past trials (historical data) and external data.
The 21st Century Cures Act, passed in the U.S. in 2016, led the Food and Drug Administration to provide guidelines for using existing data, and infrastructure is being developed to support it.
In clinical trials for rare diseases and those involving children, it is often particularly difficult to enroll a sufficient number of participants, making the effective use of historical or external data desirable \citep{limMinimizingPatientBurden2018}.

\citet{pocockCombinationRandomizedHistorical1976} focused on approaches to incorporate historical/external data, referred to hereafter as historical control data, into the control arms of current randomized controlled trials (RCTs).
\citet{spiegelhalterBayesianApproachesClinical2004} classified six types of relationships between historical and current controls for the parameter of interest, assuming that the historical control data are used following a Bayesian approach.
The meta-analytic approach \citep{neuenschwanderSummarizingHistoricalInformation2010}, which has been used in actual clinical trials \citep{baetenAntiinterleukin17AMonoclonalAntibody2013}, assumes an `` exchangeable'' relationship, in which the parameters of the historical and current controls are exchangeable.
As the meta-analytic approach is based on a random-effects meta-analysis framework, each parameter of the historical or current control is shrunk in the direction of the overall mean, depending on the heterogeneity of the historical and current control data \citep{roverBoundsWeightExternal2021}.
This shrinkage works well when all the historical control data is homogeneous with the current control data, but it may not be effective in cases involving conflicts between historical and current data.

For example, if a small number of heterogeneous historical control data are included in the meta-analytic approach, the parameter of the current control will be shrunk toward the overall mean, which is influenced by the heterogeneous historical control data. 
Consequently, the degree of borrowing from the historical control data would decrease in response to the detection of large heterogeneity \citep{ohigashiPotentialBiasModels}.
In cases where there are both homogeneous and heterogeneous historical control data, it is desirable to ignore the heterogeneous historical control data and only borrow the homogeneous historical control data \citep{hupfBayesianSemiparametricMetaanalyticpredictive2021}.
Several meta-analytic approaches to dealing with situations involving conflict have been proposed, such as the robust meta-analytic predictive (MAP) prior method \citep{schmidliRobustMetaanalyticpredictivePriors2014}, the Dirichlet process mixture(DPM) with MAP prior method \citep{hupfBayesianSemiparametricMetaanalyticpredictive2021}, which introduces a non-parametric structure to account for differences in the distribution of random effects, and the self-adapting mixture with MAP prior method \citep{yangSAMSelfadaptingMixture2023}, which avoids the effects of conflict between historical and current controls in a supervised learning framework. However, these methods are not designed to selectively borrow homogeneous historical control data.
Extensions to deal with conflict have also been introduced in methods that use assumptions other than an ``exchangeable'' relationship, but they are likewise not designed to selectively borrow homogeneous historical control data \citep{banbetaModifiedPowerPrior2019, ohigashiUsingHorseshoePrior2022, ohigashiPotentialBiasModels}.

Several factors can lead to conflict between historical and current controls, with particular attention required for systematic differences caused by unmeasured factors \citep{hupfBayesianSemiparametricMetaanalyticpredictive2021, lesaffreReviewDynamicBorrowing2024}.
Therefore, in this study, we introduce a novel parameter relationship that accounts for situations in which historical controls may conflict with current controls due to unmeasured factors, among other influences.
We then propose a nonparametric Bayesian approach that assumes this novel relationship. 
The proposed method, based on the DPM, achieves clustering according to the similarity between the historical and current control data. 
Because the DPM does not require the number of clusters to be predefined, there is no need to specify the number of conflicting historical controls in advance. 
To prioritize the resolution of conflicts between the historical and current controls over conflicts among the historical controls, we also propose a method based on the dependent DPM, which adjusts according to whether the data represents a historical or current control. 
The proposed methods possess desirable large-sample properties and ensure efficient information borrowing from homogeneous historical controls.
Moreover, these methods are flexible, as they can be implemented using the same procedure, regardless of whether the outcome data comprise aggregated study-level data or individual participant data, and can accommodate either scalars or vectors for parameters that are assumed to vary across studies.
Additionally, an index can be calculated to assess the similarity between the historical and current control data, based on the posterior distribution of the parameter of interest.
This index can be understood as ``the amount of information borrowed from the $k$th historical control.''
In existing methods, a similar interpretation can be derived from the posterior distribution of the power parameter in the modified power prior \citep{duanEvaluatingWaterQuality2006, banbetaModifiedPowerPrior2019}, but this interpretation is not derived in either the meta-analytic approach or the potential bias model \citep{hobbsHierarchicalCommensuratePower2011, ohigashiUsingHorseshoePrior2022, ohigashiPotentialBiasModels}.

\section{Methods}\label{methods}

\subsection{Dirichlet process mixture model for historical control borrowing}
Let $\theta$ denote the parameter of interest.
The Dirichlet process (DP) describes a model for a random probability distribution $G$ in a parameter space $\Theta$ \citep{fergusonBayesianAnalysisNonparametric1973, fergusonPriorDistributionsSpaces1974}.
Let $G$ be a DP with two parameters, with $M$ being the concentration parameter and $G_0$ being the base measure.
DP can be represented by the stick-breaking representation \citep{sethuramanConstructiveDefinitionDirichlet1994}:
\begin{align}
	G &= \sum^\infty_{c=1} w_c \delta_{\theta^\star_c}, 	\label{simpleDP}
	\\
	w_c &= v_c \prod_{c'<c} (1-v_{c'}) \nonumber \\
	v_c &\sim \text{Beta}(1, M), \nonumber
\end{align}
where $\delta_{\theta}$ is the Dirac measure at $\theta$ and  $\theta^\star_c \mid G_0 \overset{\text{i.i.d}}{\sim} G_0$.
One of the characteristics of DP is discreteness, which means that realizations from DP are discrete with a probability of 1.
The DPM extends the discrete DP to a continuous distribution by having realizations from the DP corresponding to latent variables.
DPM is often employed for clustering owing to its ability to form a mixture distribution without the need to specify an upper limit for the number of components \citep{lauBayesianModelbasedClustering2007}.  

\begin{figure}
	\centering
	\includegraphics[width=0.7\linewidth]{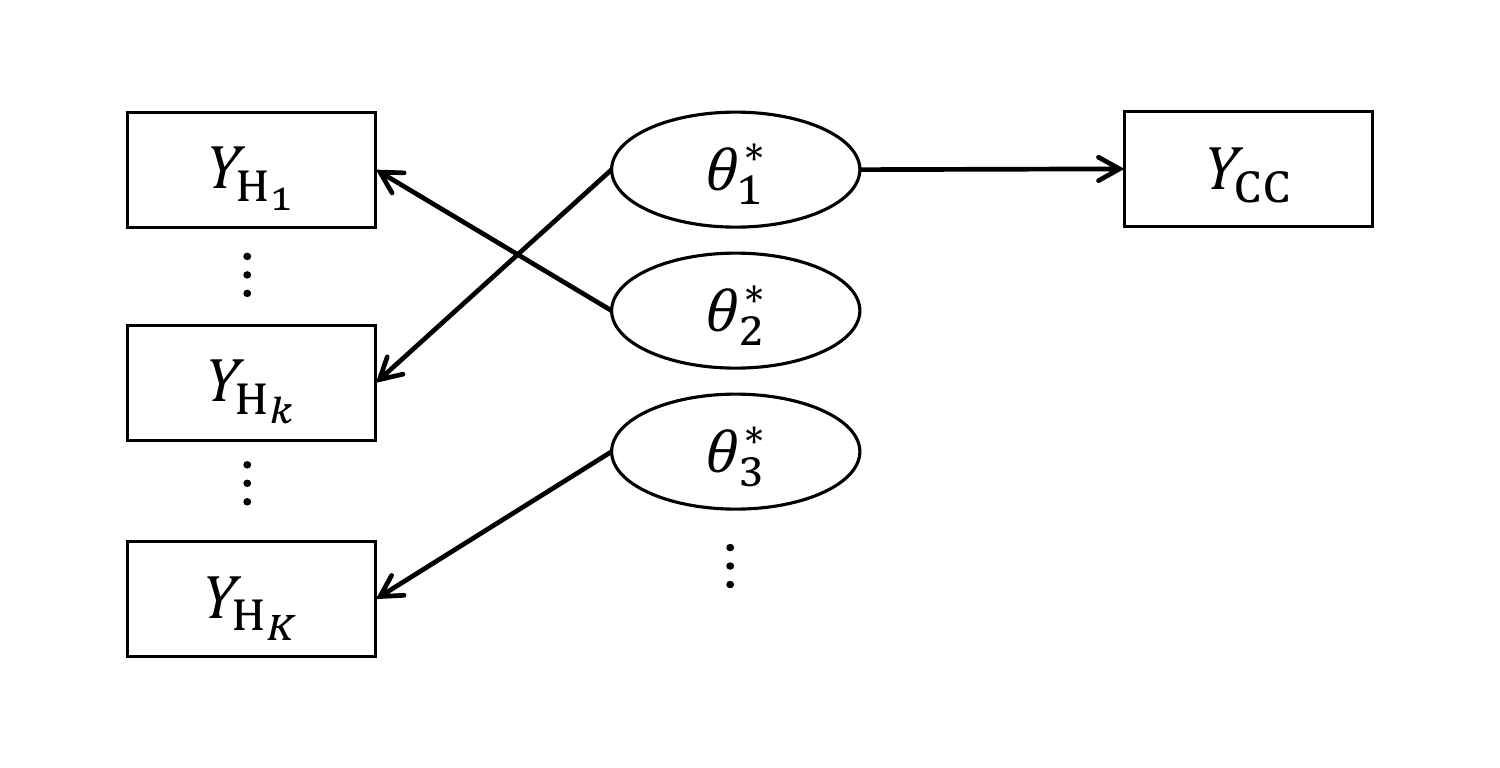}
	\caption{Potential irrelevance assumption. \label{assume}}
\end{figure}

For situations in which historical control data are used, we propose a method that utilizes DPM.
Let $\theta_{\text{H}_k}$ and $\theta_{\rm{CC}}$ denote the parameters of interest, e.g., the log-odds of response rate for a binary outcome, for the $k$th historical and current controls, respectively.
Let $y_j \left(j = \text{H}_1, \ldots, \text{H}_K, \text{CC}\right)$ denote the data, then DPM can be represented as $y_j \mid \theta_j \sim f(y_j \mid \theta_j), \theta_j \sim G, G \sim \text{DP}(M, G_0)$.
Clustering by DPM allows estimation of the number of clusters without setting an upper limit on this number.
Let $\theta_c^{\star}$ denote the realizations from DP, where the subscript $c$ is the index of discrete realization.
For example, if $\text{H}_k$ and $\text{CC}$ are assigned to the same cluster $c (=1, 2, 3, \ldots)$ during clustering, then $\theta_{\text{H}_k} = \theta_{\rm{CC}} = \theta_c^{\star}$.
This method does not correspond to the six relationship types classified by \citet{spiegelhalterBayesianApproachesClinical2004}, and it can be interpreted as a combination of two of these relationship types. In this study, we define the relationship in our proposed method as \textit{potential irrelevance}, as shown in Figure \ref{assume}.
This assumption can be viewed as a combination of the ``irrelevance'' assumption, which posits that the parameters of the historical and current controls are unrelated, and the ``equal'' assumption, which asserts that the parameters of all the historical and current controls are identical. 
As \citet{spiegelhalterBayesianApproachesClinical2004} also note that their classifications can be combined, we consider this assumption to be valid. 
It accommodates the possibility that some historical controls align with the current control, being in the same cluster, while others may be irrelevant. 
Thus, this assumption is reasonable,  and it addresses the need to accommodate conflicts arising from systematic differences driven by unmeasured factors \citep{hupfBayesianSemiparametricMetaanalyticpredictive2021, lesaffreReviewDynamicBorrowing2024}.

We can set scalars and vectors for the parameters of interest $\theta_j$.
For instance, when individual participant data with outcome $y$ and $p$-dimensional covariates $\boldsymbol{x}$ are available, and where the first component is an intercept, a linear model $y=\boldsymbol{x \beta} + \varepsilon$ may be set.
If the coefficients of the covariates are common across studies but the intercept $\beta_0$ may differ between studies, we set $\theta_j = \{ \beta_{0,j} \}$ for the DPM method.
If the coefficients, including the intercept, may differ between studies, we set $\theta_j = \{ \boldsymbol{\beta}_j \}$.
If both the coefficients and error variance $\sigma^2$ may differ between studies, we set $\theta_j = \{ \boldsymbol{\beta}_j , \sigma^2_j\}$.

\cite{hupfBayesianSemiparametricMetaanalyticpredictive2021} proposed a method for applying DPM to historical control borrowing, but it is based on a non-parametric structure to account for differences in the distribution of random effects.
Notably, when only a single historical control dataset is available, the posterior distribution of the DPM method reduces to the posterior distribution of a robust mixture prior \citep{callegaroHistoricalControlsClinical2023}.

\subsection{Dependent DPM model for historical control borrowing}
When incorporating historical controls during the design phase of a current study, it is crucial to recognize that the historical and current data may differ due to unmeasured factors \citep{lesaffreReviewDynamicBorrowing2024}.
To address this, we extend the DPM to be sensitive to unexpected differences between historical controls and current control.
We employ the dependent DP (DDP), an extension that relates predictors to DP \citep{quintanaDependentDirichletProcess2022}.
DDP is defined as an extension of the stick-breaking representation of DP.
By marginalizing DDP for all possible predictor values, we can recover the DP-distributed random measures \citep{quintanaDependentDirichletProcess2022}.

Although there are various extensions of DDP, we employ a single-atoms DDP because all the atoms $\theta_c^\star$ are linked to observed data \citep{quintanaDependentDirichletProcess2022}.
Let $s$ be a predictor related to DP, then DDP is constructed as follows:
\begin{equation*}
	G_x = \sum_{c=1}^{\infty} \left\{v_c (s) \prod_{c'<c} (1-v_{c'}(s))\right\} \delta_{\theta^\star_c}, 
\end{equation*}
where $v_c(s)$ denotes the weight that depends on $s$ and affects the mixture component $w_c$ of DP.
However, the Dirac measure $\delta_\theta$ is independent of $s$ and is common to all $s$, which is why this is known as the ``single-atoms'' approach.
Using this DDP, the prior probability model for partitioning changes according to the value of $s$. 
This is important for reflecting the implicit partitioning of the historical and current controls.
\citet{gutierrezTimeDependentBayesian2016} realized a time-dependent DP with a Markov chain structure for the time-dependent stick-breaking process, using time as a predictor, and applied it in the analysis of air quality data.

For situations in which historical controls are used, we propose a dependent DPM (DDPM) method that uses a predictor as an indicator variable to distinguish between the historical and current controls, as follows:
\begin{align*}
y_j &\sim f(y_j \mid \theta_j), \ \ j = \text{H}_1, \ldots, \text{H}_K, \text{CC} \\
\theta_j &\sim
\begin{cases}
	G_\text{H}, & j\ \in \{\text{H}_1, \ldots , \text{H}_K\} \\
	G_\text{CC}, & j=\text{CC}
\end{cases}
\\
G_\text{H} &= \sum_{c=1}^{\infty} w_c^{(\text{H})} \delta_{\theta^\star_c}, \\
G_\text{CC} &= \sum_{c=1}^{\infty} w_c^{(\text{CC})} \delta_{\theta^\star_c} 
\end{align*}
where $G_{(\cdot)}$ is a random probability measure for historical or current controls and $w_c^{(\cdot)}$ represents the weight of the stick-breaking representation for historical or current controls.
To preserve the stick-breaking structure \ref{simpleDP} at the marginal level, let $w_c^{(\cdot)}$ be the realization of $W_c^{(\cdot)}$, where
\begin{align*}
	W_1^{(\cdot)} = V_1^{(\cdot)}, \ \ W_c^{(\cdot)} = V_c^{(\cdot)} \prod_{c'<c} \left( 1-V_c^{(\cdot)} \right).
\end{align*}
Each component $W_c^{(\text{CC})}$ depends on the corresponding component of $W_c^{(\text{H})}$ as follows:
\begin{align}
	V_c^{(\text{H})} &\sim \text{Beta} (1, M) \label{DDPmarkov} \\
	V_c^{(\text{CC})} \mid V_c^{(\text{H})} &= 
	\begin{cases}
		V_c^{(\text{CC})} \sim \text{Beta} (1,M), & \text{with probability } \phi \\
		V_c^{(\text{CC})} = V_c^{(\text{H})} & \text{with probability } 1-\phi 
	\end{cases}
	\nonumber 
\end{align}
where $\phi \in [0, 1]$.
Thus, each stick-breaking component is updated with uncorrelated values from $\text{Beta}(1, M)$ with a probability $\phi$.
This construction ensures that $G_{(\cdot)}$ is the DP.
The parameter $\phi$ controls the correlation between the historical and current controls of each weight and the overall correlation between $G_\text{H}$ and $G_\text{CC}$.

Furthermore, in this study, we assign a prior distribution to the concentration parameter $M$ of DP.
This is necessary to obtain the desired statistical properties in Subsection \ref{property}.
The gamma distribution is assigned as the first choice of prior distribution for $M$, and Gibbs sampling, proposed by \citet{escobarBayesianDensityEstimation1995}, is performed.

Similar to the DPM method, the DDPM method is highly flexible, allowing the construction of a mixture distribution without the need to specify an upper limit for the number of components.
Additionally, scalars and vectors can be assigned to the parameters of interest $\theta_j$, as in the DPM method.

\subsubsection{Posterior computation}\label{posterior_comp}
To describe the Markov chain Monte Carlo algorithm, we assume for simplicity that all studies provide available summary statistics that follow $y_j \sim f(y_j \mid \theta_j)$.
Specifically, we construct a Gibbs sampler algorithm incorporating slice sampling steps, as described in \citet{walkerSamplingDirichletMixture2007}, to achieve the infinite dimensionality inherent to DDP.
We consider the augmented model given by
\begin{align*}
	f_{G_{(\cdot)}} (y_j, u_j^{(\cdot)}, z_j) = \mathds{I} \left(u_j^{(\cdot)} < w_{z_j}^{(\cdot)}\right) f(y_j \mid \theta_{z_j}),
\end{align*}
where $z$ denotes the allocation variable of $y$ and $u^{(\cdot)}$ is a uniform random variate within $(0, w_{z_j}^{(\cdot)})$.
Therefore, the augmented likelihood is expressed as:
\begin{align*}
	L_{\mathbf{V}, \mathbf{\theta}}
	 = \left\{ \prod_{j \in (\text{H}_1 ,\ldots, \text{H}_K)}
	 \mathds{I} \left(u_j^{(\text{H})} < w_{z_j}^{(\text{H})}\right)
	 f(y_j \mid \theta_{z_j})  \right\}
	  \times \mathds{I} \left(u_\text{CC}^{(\text{CC})} < w_{z_\text{CC}}^{(\text{CC})}\right)
	 f(y_\text{CC} \mid \theta_{z_\text{CC}}), \nonumber
\end{align*}
where $\mathbf{V} = \{ (v^{(\rm H)}_1, v^{(\rm H)}_2, \ldots), (v^{(\rm CC)}_1, v^{(\rm CC)}_2, \ldots)\}$ denote the infinite collection of  observations for the historical and current controls of the stick-breaking components (\ref{DDPmarkov}).

The main variables $v^{(\cdot)}_c, \theta^\star_c, z_j$, and $u_j^{(\cdot)}$ are sampled at each step using the Gibbs sampler algorithm.
Similarly, $M$ is sampled using the Gibbs sampler algorithm if a gamma distribution is assigned to the prior of $M$.
However, an independent Metropolis–Hasting sampler algorithm is necessary for $\phi$ because the posterior distribution for $\phi$ is not available in a closed form.
The algorithms for all posterior inferences and the model, applicable when individual participant data are available, are detailed in the Supplementary materials.

As a by-product of the posterior computation, we can obtain samples from the posterior distribution of the latent variable $z_j$.
The number of times $z_j = z_{j'}$ in the posterior sample corresponds to the number of times study $j$ was assigned to the same cluster as study $j'$, and is often used as an indicator to evaluate the similarity between studies \citep{mullerBayesianNonparametricData2015}.
However, our focus is on the similarity between the current control and each historical control, defining $\text{Pr}(z_\text{CC} = z_{\text{H}_k} \mid D) $ as the ``Similarity and Borrowing Index (SBI),'' which will be evaluated in Section \ref{case}.

\subsection{Statistical properties}\label{property}
The proposed methods exhibit two desirable large-sample properties. 
The proof is provided in the Supplementary materials.
First, when the number of historical controls is large, we obtain the following consistency result.

\begin{theorem}
The proposed methods achieve consistency in the number of clusters when historical and current control data are generated from a finite mixture.
\end{theorem}

As demonstrated in Theorem 1 in \citet{ascolaniClusteringConsistencyDirichlet2023}, if the prior distribution for the concentration parameter $M$ satisfies certain conditions, the posterior of the number of clusters is consistent with the true number of clusters.
In other words, it assigns historical and current controls to the same cluster if they are in the same cluster in the actual data-generating structure and appropriately categorizes historical controls in different clusters as irrelevant.
Second, when the sample size of each study is large, we obtain the following consistency result.

\begin{theorem}
When the sample size of each study is large, the posterior distributions of $\theta_{\text{CC}}$ and $\theta_{\text{H}_k}$ converge to the same true value if the current control and the $k$th historical control are congruent, and to distinct values if the current control and the $k$th historical control are incongruent.
\end{theorem}

This theorem is derived from the properties of posterior consistency and the posterior distribution of DP. 
Consequently, when the sample size is large, the results align with the estimates obtained from RCT data alone.

\section{Simulation study}\label{sim}

\subsection{For summary statistics}

\subsubsection{Settings}\label{simsum}
We evaluate the performance of the proposed and existing methods using simulation studies.
We consider a current trial, as well as eight historical trials, with 20 and 40 participants for the current control and treatment groups, respectively, and 60 participants for the historical controls.
The outcome is binary, and the number of responses is generated from a binomial distribution.
In the first scenario, there is no between-trial heterogeneity.
The common response rate of the historical and current controls, $\pi_j$, is 50\%.
Therefore, either assumption is acceptable for use in this scenario.
In the second scenario, the between-trial heterogeneity is large.
The log odds of the response rates of the historical and current controls follow a normal distribution with mean 0 and variance $0.5^2$, $\text{logit}\left(\pi_j\right) \sim \text{N}(0, 0.5^2)$.
\citet{spiegelhalterBayesianApproachesClinical2004} and \citet{neuenschwanderSummarizingHistoricalInformation2010} suggested that in practice, between-trial heterogeneity, in terms of standard deviation (SD), often lies between 0.1 and 0.5 on the log-odds scale. 
Approximating this heterogeneity, the SD of the response rate yields 12.5\%. 
Thus, meta-analytic approaches that rely on the ``exchangeable'' assumption are appropriate in this scenario, as they directly model between-trial variance.

In the third, fourth, and fifth scenarios, to assess the influence of heterogeneous historical controls, several historical controls follow a distribution different from that of the current control.
In the third scenario, the number of heterogeneous historical controls is two.
The response rate of the homogeneous historical and current controls is 50\% and that of the heterogeneous historical and current controls is 20\%.
In the fourth scenario, the number of heterogeneous historical controls is four, and the response rates are the same as in the third scenario.
In the fifth scenario, the number of heterogeneous historical controls increases to eight, while the response rates remain unchanged.
In scenarios with the null hypothesis, the response rate to the current treatment, $\pi_\text{CT}$, is 50\%.
In scenarios with the alternative hypothesis, $\pi_\text{CT}$ is 74.52\%, and the power reaches 50\% based on the chi-square test for the current trial, assuming a response rate of 50\% for the current control.
Because the overall means are influenced by the heterogeneous historical controls in these scenarios, methods based on the ``exchangeable'' assumption are not appropriate.

We compare the DPM and DDPM using the following methods: 
(1) Current data (CD) analysis, which uses only current RCT data with $\pi_\text{CC}, \pi_\text{CT} \sim \text{Beta}(0.5, 0.5)$;
(2) Pooled data (PD) analysis, which pools current and historical controls with $\pi_\text{(CC+H)} \sim \text{Beta}(0.5, 0.5)$; 
(3) A MAP prior method \citep{schmidliRobustMetaanalyticpredictivePriors2014} that generates a predictive prior for $\pi_\text{CC}$ based on historical controls, with a non-informative prior, $\text{N}(0, 10^2)$, for the overall mean and a weakly informative prior for the between-trial standard deviation, with a half-normal distribution with variance 1, using the \texttt{RBesT} package \citep{weberApplyingMetaAnalyticPredictivePriors2021};
(4) A DPM with MAP prior (DPM-MAP) method \citep{hupfBayesianSemiparametricMetaanalyticpredictive2021} that generates a predictive prior similar to the MAP method; however, it assumes that the between-trial variance follows DPM, whose base distribution is half-normal distribution with variance 1 and a fixed concentration parameter 1;
(5) A self-adapting mixture prior to the MAP prior (SAM-MAP) method \citep{yangSAMSelfadaptingMixture2023}, which is an empirical Bayesian approach. 
In the SAM-MAP method, the clinically significant difference parameter is 0.1, which is often used as the non-inferiority margin for non-inferiority trials in which the response variable is binary, using the \texttt{SAMprior} package.
In the DPM and DDPM methods we assign a beta distribution $\text{Beta}(0.5, 0.5)$ for the base distribution.
For the concentration parameter $M$, we assign a gamma prior, $M \sim \text{Gamma}(1, \text{scale=}5)$.
In the Supplementary materials, we show the results for the other hyperparameter settings for $M$.
In DDPM, we assign a beta distribution $\phi \sim \text{Beta}(2, 2)$ for $\phi$, following \citet{gutierrezTimeDependentBayesian2016}.

We calculate the bias of the mean of the posterior distribution of the treatment effect as $\pi_\text{CT} - \pi_\text{CC}$.
We also calculate the frequentist type I error rate and power when the treatment effect is considered statistically significant and the posterior probability $\text{Pr}(\pi_\text{CT} - \pi_\text{CC} > 0 \mid D)$ is greater than 0.975.
To assess the average number of historical controls incorporated by each method, except for CD and PD, we also calculate the mean of the effective historical sample size (EHSS) \citep{hobbsAdaptiveAdjustmentRandomization2013, wiesenfarthQuantificationPriorImpact2020, bennettNovelEquivalenceProbability2021}.
The EHSS is the posterior ESS minus the sample size of the current control, and the posterior ESS is calculated using the ELIR method \citep{kaizerDiscussionPredictivelyConsistent2020}.
The posterior distribution of $\pi_\text{CC}$ is approximated by three beta distributions using the \texttt{mixfit} function in the \texttt{RBesT} package, and the \texttt{ess} function is used to obtain the posterior ESS using the ELIR method.
We also calculate the root mean square error, mean posterior SD, and coverage probability of the posterior distribution of the treatment effect $\pi_\text{CT} - \pi_\text{CC}$.

We use Stan via the \texttt{cmdstanr} version 0.4.0 package for the DPM-MAP method and \texttt{R} version 4.2.3 for Linux on the supercomputer system of the Institute of Statistical Mathematics, Tokyo, Japan. 
We generate more than 10000 datasets and present the results of 10000 iterations that reach the convergence criteria described above.
The Supplementary materials contain all the results.
 
\subsubsection{Results}
Table \ref{res_sum} shows the bias, frequentist type I error rate, power, and mean of the EHSS.
In scenario 1, the biases for all methods are near zero, and the type I error rates for all methods remain approximately 2.5\%.
The power for DPM and DDPM are higher than those for the three methods using MAP in scenario 1, in which no between-trial heterogeneity is observed (although it does not approach PD).
In scenario 2, in which the ``exchangeable'' assumption holds, the type I error rates for DPM and DDPM increase, whereas those for the meta-analytic approaches remain close to 2.5\%. 
The power for MAP and DPM-MAP exceeds that for CD.
In scenarios 3, 4, and 5, in which the ``exchangeable'' assumption fails due to existing heterogeneous historical controls, the biases for SAM-MAP remain small.
In scenarios 3 and 4, the biases for DPM and DDPM are smaller than those for MAP and DPM-MAP. 
In contrast, in scenario 5, only the bias for DDPM is smaller than those for MAP and DPM-MAP.
The type I error rates for DPM and DDPM significantly increase in scenarios 4 and 5 because the biases of DPM and DDPM are increased.

The power for SAM-MAP in scenarios 2 to 5 is nearly identical to that for CD. 
This shows that SAM-MAP does not borrow information from the historical controls.
This is supported by the mean EHSS results for SAM-MAP.
In DPM and DDPM, the differences in the mean EHSS between scenarios 1 and 4 are almost equal to the number of participants in the heterogeneous historical controls (60 $\times$ 4); thus, the heterogeneous historical controls can be identified.
Conversely, the result observed in scenario 4 cannot be replicated in scenario 5 because the small sample size precludes the application of Theorem 2.

\begin{table}[tbp]
	\caption{Bias of the posterior mean of the posterior distribution of treatment effect $\pi_\text{CT} - \pi_\text{CC} (\%)$, type I error rate (\%), power (\%), and mean effective historical sample size in the simulation study for summary statistics.\label{res_sum}}
	\centering
	\begin{tabular}{lrrrrr}
		\toprule
		Method  & Sce 1 & Sce 2 & Sce 3 & Sce 4 & Sce 5 \\
		\midrule
		\multicolumn{6}{l}{\textbf{Bias}} \\
		CD      & $-$0.78 & $-$1.58 & $-$0.85 & $-$1.51 & $-$0.17 \\
		PD      & $-$0.63 & $-$1.63 & 6.50  & 13.71 & 28.14 \\
		MAP     & $-$0.67 & $-$1.56 & 1.81  & 2.87  & 17.28 \\
		DPM-MAP & $-$0.69 & $-$1.56 & 0.91  & 2.98  & 9.65  \\
		SAM-MAP & $-$1.61 & $-$2.36 & $-$0.59 & $-$1.29 & $-$0.15 \\
		DPM     & $-$0.65 & $-$1.51 & 0.74  & 1.42  & 12.07 \\
		DDPM    & $-$0.66 & $-$1.52 & 0.89  & 1.25  & 9.47  \\
		\midrule
		\multicolumn{6}{l}{\textbf{Type I error rate (\%)}} \\
		CD      & 2.33  & 2.60  & 2.61  & 2.53  & 2.84  \\
		PD      & 2.65  & 13.86 & 14.81 & 44.44 & 96.66 \\
		MAP     & 1.65  & 2.80  & 3.17  & 4.17  & 36.91 \\
		DPM-MAP & 1.49  & 2.59  & 2.45  & 4.34  & 15.11 \\
		SAM-MAP & 1.31  & 2.07  & 2.53  & 2.53  & 2.84  \\
		DPM     & 2.24  & 3.90  & 2.60  & 3.44  & 25.80 \\
		DDPM    & 2.14  & 3.83  & 2.85  & 3.54  & 19.20 \\	
		\midrule
		\multicolumn{6}{l}{\textbf{Power (\%)}} \\
		CD      & 47.2  & 46.5  & 46.9  & 44.3  & 49.1  \\
		PD      & 87.9  & 74.9  & 98.1  & 99.9  & 100.0 \\
		MAP     & 78.4  & 57.0  & 63.9  & 64.2  & 95.2  \\
		DPM-MAP & 70.4  & 55.6  & 64.9  & 64.2  & 69.5  \\
		SAM-MAP & 55.1  & 47.2  & 49.6  & 44.7  & 49.1  \\
		DPM     & 84.0  & 60.2  & 78.7  & 75.6  & 63.8  \\
		DDPM    & 82.4  & 59.9  & 78.2  & 75.5  & 59.4  \\
		\midrule
		\multicolumn{6}{l}{\textbf{Mean EHSS}} \\
		MAP     & 123.2 & 21.8  & 10.4  & 7.1   & 65.0  \\
		DPM-MAP & 62.8  & 17.9  & 16.5  & 7.5   & 20.1  \\
		SAM-MAP & 36.6  & 6.2   & 2.5   & 1.5   & 1.3   \\
		DPM     & 373.9 & 115.8 & 197.7 & 129.3 & 192.7 \\
		DDPM    & 394.5 & 130.0 & 210.2 & 143.5 & 174.2  \\
		\bottomrule
	\end{tabular}	
	\begin{tablenotes}
		\item Abbreviations: CD, current data analysis; PD, pooled data analysis; MAP, meta-analytic predictive prior method; DPM-MAP, Dirichlet process mixture-meta-analytic predictive prior method; SAM-MAP, self-adapting mixture-meta-analytic predictive prior method; DPM, Dirichlet process mixture method; DDPM, dependent Dirichlet process mixture method; EHSS, effective historical sample size.
		\item Note: The five scenarios are defined as follows: Sce 1, no between-trial heterogeneity; Sce 2, large between-trial heterogeneity; Sce 3, two heterogeneous historical controls exist; Sce 4, four heterogeneous historical controls exist; Sce 5, eight heterogeneous historical controls exist.
	\end{tablenotes}
\end{table}

\subsection{For individual participant data}

\subsubsection{Settings}
We evaluate the performance of the proposed and existing methods using simulation studies in situations where individual participant data is available as historical and current data.
We use the following settings based on a case of Alzheimer's disease, which is the motivational data of \citet{qiIncorporatingHistoricalControl2022}.
We consider a current trial and five historical trials with 60 participants in the current control and treatment groups and 100 participants in the historical controls.
We generate $\text{age}_{j,i}$ from $\text{N}(\mu_{\text{age,}j}, 8^2)$, where $\mu_{\text{age},j}$ is drawn from $\text{Unif}(71, 77)$ to represent the mean age in the $j$th study.
Similarly, $\text{sex}_{j,i}$ is generated from $\text{Bernoulli}(\pi_{\text{F,}j})$, where $\pi_{\text{F,}j}$ is drawn from $\text{Unif}(0.5, 0.6)$  to represent the proportion of females in the $j$th study.

We consider four scenarios for the mean baseline Alzheimer’s Disease Assessment Scale - Cognitive (ADAS-cog) in the $j$th study.
In the first scenario, there is no between-trial heterogeneity; the baseline ADAS-cog is generated from a common normal distribution with a mean of $\mu_{\text{base},j} = 24$.
In the second scenario, there is large between-trial heterogeneity, with $\mu_{\text{base},j}$ generated from $\text{N}(24, 3^2)$.
Thus, meta-analytic approaches that rely on the ``exchangeable'' assumption are appropriate in this scenario, as they directly model between-trial variance.
In the third and fourth scenarios some of the historical controls follow distributions distinct from the current control, exploring the impact of heterogeneous historical controls. 
In the third scenario, one historical control is heterogeneous, with $\mu_{\text{base},j}$ set to 30, while the homogeneous historical and current controls have $\mu_{\text{base},j} = 24$.
This setup mirrors the motivational data in \cite{qiIncorporatingHistoricalControl2022}. 
In the fourth scenario, two historical controls are heterogeneous, maintaining the same means as in the third scenario. 
Because the overall means are influenced by the heterogeneous historical controls in scenarios 3 and 4, methods based on the ``exchangeable'' assumption are not appropriate.
For each scenario, $\text{base}_{j,i}$ is generated from $\text{N}(\mu_{\text{base},j}, 8^2)$.
 
The primary outcome $Y_{j,i}$ is a change in ADAS-cog, a score of cognitive function, after 12 months.
We generate this primary outcome as follows:
\begin{align*}
	Y_{j,i} = 5 + 0.2\ (\text{age}_{j,i} - 50) + 1.0\ \text{sex}_{j,i} + 1.0\ \text{base}_{j,i} + \gamma\text{TRT}_{j,i} + \varepsilon_{j,i}, \ \ \varepsilon_{j,i} \sim \text{N}(0, 1),
\end{align*}
where $\text{TRT}_{j,i}$ is a treatment indicator equal to zero for the control group and one for the treatment group.

We compare the DPM and DDPM using the following methods:
(1) CD analysis, which only uses current RCT data with $\boldsymbol{\beta} \sim \text{N}(0, 1000^2 \boldmath{I})$ and $\sigma^2 \sim \text{invGamma}(0.01, 0.01)$;
(2) PD analysis, which pools current and historical controls;
(3) An exchangeable (EX) method \citep{hanCovariateadjustedBorrowingHistorical2017} that assumes partial exchangeability for the intercept of the $j$th study $\beta_{0, j}$, which is applicable when the current and historical controls are exchangeable after accounting for the baseline covariates, with $\beta_{0, \text{H}_1}, \ldots \beta_{0, \text{H}_5}, \beta_{0, \text{C}} \sim \text{N}(\mu, \tau^2), \mu \sim \text{N}(0, 100^2), \tau \sim \text{halfN}(5)$.
In DPM and DDPM, we assume that the coefficients of the covariates differ between each study and assign the non-informative prior, $\boldsymbol{\beta} \sim \text{N}(0, 1000^2)$ for the base distribution.
For common error variance, we assign the non-informative prior $\sigma^2 \sim \text{invGamma}(0.01, 0.01)$.
For the other parameters in DPM and DDPM, we assign prior distributions, as in Subsection \ref{simsum}, $M \sim \text{Gamma}(1, \text{scale=}5)$ and $\phi \sim \text{Beta}(2, 2)$.
To assess the performance of each method in situations where the between-trial heterogeneity or presence of heterogeneous historical controls arise from an unmeasured factor, we treat the baseline ADAS-cog as an unmeasured factor and exclude it from the analytical model.

We calculate the bias of the posterior mean of the posterior distribution of treatment effect $\beta_\text{TRT}$.
We also calculate the frequentist type I error rate and power when the treatment effect is considered statistically significant and the posterior probability $\text{Pr}(\beta_\text{TRT} < 0 \mid D)$ is greater than 0.975.
To assess the average number of historical controls incorporated by DPM and DDPM, we calculate the mean of the posterior mean of the number of historical controls assigned to the same cluster as the current control.
We also calculate the root mean square error, mean posterior SD, and coverage probability of the posterior distribution of treatment effect.

We use Stan via the \texttt{cmdstanr} package (version 0.4.0) for the EX method, and \texttt{R} version 4.2.3 for Linux on the supercomputer system of the Institute of Statistical Mathematics, Tokyo, Japan. 
We generate more than 2000 datasets and present the results of 2000 iterations that reach the convergence criteria described above.
Results not included in the main text are included in the Supplementary materials.

\subsubsection{Results}
Table \ref{res_ipd} lists the bias, frequentist type I error rate, power, and mean of the number of historical controls in the same cluster.

In scenario 1, the biases for all methods are near zero, and the type I error rates for all methods remain below approximately 2.5\%.
Both DPM and DDPM achieve power comparable to PD and borrow all the historical controls.
This is supported by the posterior mean of the number of historical controls assigned to the same cluster as the current control.
However, the power for EX in scenario 1 does not reach PD, indicating that EX does not actively borrow from the historical controls, even in the absence of between-trial heterogeneity.
In scenario 2, in which the ``exchangeable'' assumption holds, the type I error rates for DPM and DDPM significantly increase, while that for EX remains below approximately 2.5\% due to its alignment with the true model.
In scenarios 3 and 4, in which the ``exchangeable'' assumption fails due to existing heterogeneous historical controls, the biases and type I error rates for EX are larger than those for DPM and DDPM.
The posterior mean of the number of historical controls shows that DPM and DDPM can identify the heterogeneous historical controls in scenarios 3 and 4.

\begin{table}[tbp]
	\caption{Bias of the posterior mean of the posterior distribution of treatment effect $\beta_\text{TRT}$, type I error rate (\%), power (\%), and mean of the posterior mean of the number of historical controls assigned to the same cluster  in the simulation study for individual participant data.\label{res_ipd}}
	\centering
	\begin{tabular}{lrrrr}
		\toprule
		Method  & Sce 1 & Sce 2 & Sce 3 & Sce 4 \\
		\midrule
		\multicolumn{5}{l}{\textbf{Bias}} \\
		CD   & 0.06  & $-$0.03 & $-$0.03 & $-$0.07 \\
		PD   & 0.03  & $-$0.15 & $-$1.08 & $-$2.18 \\
		EX   & 0.03  & $-$0.04 & $-$0.17 & $-$0.26 \\
		DPM  & 0.03  & $-$0.09 & $-$0.08 & $-$0.05 \\
		DDPM & 0.03  & $-$0.05 & $-$0.05 & $-$0.05 \\
		\midrule
		\multicolumn{5}{l}{\textbf{Type I error rate (\%)}} \\
		CD   & 2.70  & 2.65  & 2.35  & 2.45  \\
		PD   & 2.20  & 22.60 & 14.80 & 44.30 \\
		EX   & 1.50  & 2.70  & 2.55  & 3.10  \\
		DPM  & 2.00  & 12.95 & 4.00  & 2.25  \\
		DDPM & 2.10  & 10.95 & 3.40  & 2.05  \\
		\midrule
		\multicolumn{5}{l}{\textbf{Power (\%)}} \\
		CD   & 48.8 & 51.3 & 53.5 & 54.9 \\
		PD   & 76.9 & 62.1 & 95.7 & 99.7 \\
		EX   & 65.4 & 55.3 & 60.8 & 61.4 \\
		DPM  & 77.0 & 64.7 & 78.6 & 76.8 \\
		DDPM & 76.5 & 64.5 & 77.5 & 77.0 \\
		\midrule
		\multicolumn{5}{l}{\textbf{Mean of no. HCs in same cluster}} \\
		DPM  & 5.00  & 3.41  & 4.10  & 3.02  \\
		DDPM & 4.98  & 3.08  & 4.06  & 3.01 \\
		\bottomrule
	\end{tabular}	
	\begin{tablenotes}
		\item Abbreviations: CD, current data analysis; PD, pooled data analysis; EX, exchangeable method that assumes partial exchangeability for the intercept; DPM, Dirichlet process mixture method; DDPM, dependent Dirichlet process mixture method; no. HCs in same cluster, posterior mean of the number of historical controls in the same cluster.
		\item Note: The four scenarios are defined as follows: Sce 1, no between-trial heterogeneity; Sce 2, large between-trial heterogeneity; Sce 3, one heterogeneous historical control exists; Sce 4, two heterogeneous historical controls exist.		
	\end{tablenotes}
\end{table}

\section{Case study}\label{case}
In this section, we describe the application of the proposed and existing methods to a clinical trial.
This example is from a phase II proof-of-concept trial for the treatment of ankylosing spondylitis (AS) using secukinumab \cite{baetenAntiinterleukin17AMonoclonalAntibody2013}.
The outcome is binary (achievement of a 20\% response according to the SpondyloArthritis International Society criteria for improvement). 
Table \ref{Case_Data} presents these data.
To analyze the data, we use the same methods and settings as those used in the simulation study.
Analysis of the AS study is presented as Case 1.
As in the simulation study, to investigate model behavior when heterogeneous historical controls exist, a case in which the number of responses in Study 3 is modified from 19 to 31, defined as Case 2, is analyzed.
Hence, the response rate for Study 3 in Case 2 is 60.8\%.
\begin{table}[tbp]
	\caption{Observed sample sizes, number of responses, and response rate in ankylosing spondylitis examples for Cases 1 (original dataset) and 2 (modified dataset).\label{Case_Data}}
	\centering
	\begin{tabular}{llrrrrrrrrrrrr}
		\toprule
		&& \multicolumn{9}{c}{Historical controls} && \multicolumn{2}{c}{Current trial} \\
		\cmidrule{3-11}\cmidrule{13-14}
		Case &Trial    & 1    & 2    & 3    & 4    & 5    & 6  & 7    & 8    & (Sum) && CC   & CT   \\
		\midrule
		1&Response & 23   & 12   & 19   & 9    & 39   & 6  & 9    & 10  & 127 && 1    & 14   \\
		&N        & 107  & 44   & 51   & 39   & 139  & 20 & 78   & 35 & 513   && 6    & 23   \\
		&\%       & 21.5 & 27.3 & 37.3 & 23.1 & 28.1 & 30.0 & 11.5 & 28.6 & 24.8&& 16.7 & 60.9 \\
		\midrule
		2 &Response & 23   & 12   & 31   & 9    & 39   & 6  & 9    & 10  & 149  && 1    & 14   \\
		&N        & 107  & 44   & 51   & 39   & 139  & 20 & 78   & 35   & 513 && 6    & 23   \\
		&\%       & 21.5 & 27.3 & 60.8 & 23.1 & 28.1 & 30.0 & 11.5 & 28.6 & 29.0 && 16.7 & 60.9 \\
		\bottomrule
	\end{tabular}
\end{table}

Table \ref{Case_Res} summarizes the posterior distributions of the treatment effect (the difference between the response rates of the secukinumab and placebo groups) estimated using the above methods.
The 95\% CIs for the treatment effect do not contain zero for any method.
\begin{table}[tbp]
	\caption{Summary of the posterior distributions of the treatment effect, $\pi_\text{CT}-\pi_\text{CC} (\%)$, for Cases 1 (original dataset) and 2 (modified dataset).\label{Case_Res}}
	\centering
	\begin{tabular}{lrrcrrrcr}
		\toprule
		\multicolumn{1}{c}{ } & \multicolumn{4}{c}{Case 1} & \multicolumn{4}{c}{Case 2} \\
		\cmidrule(l{3pt}r{3pt}){2-5} \cmidrule(l{3pt}r{3pt}){6-9}
		Method & Mean & SD & 95\%CI & EHSS & Mean & SD & 95\%CI & EHSS\\
		\midrule
		CD      & 39.1 & 17.4 & (0.3, 68.5)  &      & 39.2 & 17.4 & (0.7, 68.5)  &      \\
		PD      & 35.7 & 9.9  & (15.5, 54.1) &      & 33.4 & 10.0 & (13.2, 52.1) &      \\
		MAP     & 36.5 & 11.9 & (12.2, 58.8) & 37.1  & 36.4 & 14.2 & (7.1, 62.3)  & 10.9  \\
		DPM-MAP & 36.4 & 12.4 & (11.3, 60.0)   & 26.4  & 36.9 & 13.4 & (9.5, 61.7)  & 15.2  \\
		SAM-MAP & 37.9 & 15.3 & (4.8, 65.7)  & 8.5   & 37.7 & 16.1 & (3.1, 66.0)    & 3.0   \\
		DPM     & 35.8 & 11.2 & (13.6, 57.6) & 173.9 & 36.6 & 12.2 & (11.2, 59.2) & 138.7 \\
		DDPM    & 36.1 & 11.2 & (13.9, 58.1) & 220.5 & 36.8 & 12.3 & (10.7, 59.3) & 163.4 \\
		\bottomrule
	\end{tabular}
	\begin{tablenotes}
		\item Abbreviations: SD, standard deviation; CI, credible interval; CD, current data analysis; PD, pooled data analysis; MAP, meta-analytic predictive prior method; DPM-MAP, Dirichlet process mixture-meta-analytic predictive prior method; SAM-MAP, self-adapting mixture-meta-analytic predictive prior method; DPM, Dirichlet process mixture method; DDPM, dependent Dirichlet process mixture method.
	\end{tablenotes}
\end{table}
In Case 1, the posterior means of MAP, DPM-MAP, DPM, and DDPM are close to those of PD; however, the posterior SDs of DPM and DDPM are smaller than those of MAP and DPM-MAP.
The posterior SD of DPM-MAP is larger than the posterior SD of MAP. However, this is reversed in Case 2.
This is because DPM-MAP can avoid the influence of the heterogeneous historical control by estimating the larger variance component.
In Case 2, the posterior SDs of DPM and DDPM are smaller than those of MAP and DPM-MAP.
The EHSS for SAM-MAP is extremely small in both cases and does not borrow information from the historical controls.

As noted in Subsection \ref{posterior_comp}, the proposed methods enable calculation of the frequency with which each study is assigned to the same cluster.
Table \ref{Case_Res_similarity} provides a posterior summary of the SBI, which is the proportion of the cluster membership variables of the historical controls that are assigned to the same cluster as the current control in the posterior sampling.
In Case 1, the response rate for $\text{H}_7$ is lower than those of the other historical controls, leading to a slightly lower SBI owing to its lower similarity.
In Case 2, the SBI for $\text{H}_3$, which was modified to exhibit a higher response rate, is notably small, demonstrating the capacity of the proposed methods to detect heterogeneous historical controls.
In both cases, the posterior probability for DDPM is higher than that for DPM, which is consistent with the EHSS results in Table \ref{Case_Res}.
 The difference in the SBI between Cases 1 and 2 arises because the number of studies with response rates homogeneous with the current control decreased from seven to six. 
 This reduction lowers the accuracy of parameter estimation for the relevant cluster, which in turn impacts clustering accuracy and decreases the proportion of cases in the same cluster. 
\begin{table}[tbp]
	\caption{Posterior proportion $(\%)$ of the cluster membership variable for the historical control $z_{\text{H}_k}$ allocated to the same cluster as the current control, i.e., $\text{Pr}(z_\text{CC} = z_{\text{H}_k} \mid D) $, for Cases 1 (original dataset) and 2 (modified dataset).\label{Case_Res_similarity}}
	\centering
	\begin{tabular}{lcccccccc}
		\toprule
		Method & $\text{H}_1$ & $\text{H}_2$ & $\text{H}_3$ & $\text{H}_4$ & $\text{H}_5$ & $\text{H}_6$ & $\text{H}_7$ & $\text{H}_8$ \\
		\midrule
		\multicolumn{9}{l}{\textbf{Case 1}} \\
		DPM    & 75.8         & 76.6         & 71.4         & 76.3         & 76.6         & 75.6         & 49.6         & 76.2         \\
		DDPM   & 80.3         & 80.4         & 77.9         & 80.4         & 80.4         & 80.0         & 64.9         & 80.3         \\
		\midrule
		\multicolumn{9}{l}{\textbf{Case 2}} \\
		DPM    & 70.3         & 69.7         & 1.6          & 69.8         & 68.9         & 68.2         & 45.0         & 69.4         \\
		DDPM   & 73.5         & 73.3         & 2.1          & 73.3         & 72.7         & 72.1         & 55.1         & 73.1 \\        
		\bottomrule
	\end{tabular}
	\begin{tablenotes}
		\item Abbreviations: DPM, Dirichlet process mixture method; DDPM, dependent Dirichlet process mixture method.
	\end{tablenotes}
\end{table}

\section{Discussion}\label{dis}
In this study, we propose methods that incorporate historical controls based on nonparametric Bayesian approach.
When the ``exchangeability'' assumption fails due to heterogeneous historical controls, the proposed methods mitigates their influence.
This functionality is particularly noticeable with DDPM compared to DPM in cases with a finite sample size.
Such scenarios often arise owing to unmeasured factors, and because these factors cannot be identified before obtaining the current data, understanding the operating characteristics in these situations is crucial for choosing an appropriate dynamic borrowing method \citep{hupfBayesianSemiparametricMetaanalyticpredictive2021, lesaffreReviewDynamicBorrowing2024}.

The proposed methods are adaptable to various data formats because they process each dataset uniformly, whether the outcome data comprise aggregated study-level data or individual participant data.
This is novel because existing methods do not adjust for individual participant covariates using the same procedure as that used for aggregated study-level data.
Additionally, the proposed SBI, which is easily interpretable as a measure of the similarity between the current control and each historical control, can be calculated using the proposed methods.
While it is possible to quantitatively assess whether the current control is similar to each historical control using the posterior distribution with the power prior, this assessment is not feasible when using the meta-analytic approach.
Although the power prior has been extended to handle various situations, there are cases where it cannot be applied, depending on the outcome type, the number of historical controls, and the availability of individual participant data \citep{vanrosmalenIncludingHistoricalData2018}.
The proposed methods are novel because they can quantitatively evaluate similarity while incorporating historical controls into a unified framework suitable for various situations.

In contrast to empirical Bayesian methods, the proposed methods are based on a fully Bayesian procedure and demonstrate consistency \citep{robertBayesianChoiceDecisiontheoretic2007}.
SAM-MAP is an empirical Bayesian approach.
When applying this method, careful consideration of the tuning parameters is necessary during implementation.
In this study, the commonly used non-inferiority margin served as the comparator. 
However, if there is a large amount of between-trial heterogeneity or if heterogeneous historical controls exist, SAM-MAP refrains from borrowing historical controls, making it difficult to achieve the desired dynamic borrowing.

When the response variable comprises other types of endpoints, such as time-to-event data, a fully parametric model can be constructed to perform the same procedure.
However, in cases such as inference based on the partial likelihood of a Cox proportional hazards model, the procedure may not be directly applicable, and further research is necessary.

\section*{Acknowledgement}
This study was performed under the Cooperative Use Registration (2023-ISMCRP-0003). The computations in this study were performed at the Institute of Statistical Mathematics, Tokyo, Japan. We thank Editage [http://www.editage.com] for editing and reviewing the manuscript for English language.
We would like to express our gratitude to Dr. Luis Gutiérrez for providing the code for the method proposed in Gutiérrez et al. (2016).

\section*{Funding Statement}
This work was supported by JSPS KAKENHI (grant numbers JP23K19969 and 24K20739 to TO and JP23K11015 to TO and TS).

\section*{Conflict of interest}
The authors declare no conflicts of interest.

\section*{Data Availability Statement}
The AS trial data used in this study were available from \citet{baetenAntiinterleukin17AMonoclonalAntibody2013}.

\section*{Supporting data}
The Supplementary materials are available online. 
Code for implementing the proposed methods is available as the R package \texttt{ddp4hc} at  GitHub: \url{https://github.com/tom-ohigashi/ddp4hc}.

\end{document}

\title{Supplementary Materials for ``Nonparametric Bayesian method for dynamic borrowing of historical control data''}
\author[1]{Tomohiro Ohigashi}
\author[2]{Kazushi Maruo}
\author[1]{Takashi Sozu}
\author[2]{Masahiko Gosho}
\affil[1]{Department of Information and Computer Technology, Faculty of Engineering, Tokyo University of Science, Tokyo, Japan}
\affil[2]{Department of Biostatistics, Institute of Medicine, University of Tsukuba, Tsukuba, Japan}
\date{\empty}
\date{\empty}
\maketitle

\section{Step-by-step Sampling Procedures for Dirichlet process mixture method}\label{MCMC_DPM}
For simplicity, we show the posterior sampling procedure when the outcome data $y$ are aggregated study-level data.
We construct a Gibbs sampler algorithm with slice sampling proposed by \citet{walkerSamplingDirichletMixture2007}.
We consider an augmented model for study $j \left(= \text{H}_1, \ldots, \text{H}_K, \text{CC}\right)$ given by
\begin{align*}
f (y_j , u_j \mid \boldsymbol{w, \theta}) = \sum_{c=1}^{\infty} w_c p(u_j \mid w_c) p(y_j \mid \theta_c),
\end{align*}
where $u_j$ is a uniform random variable on $(0, w_c)$.
Let $z_j$ is the allocation variable of $y_j$, the augmented model is given by
\begin{align*}
f (y_j, z_j = c , u_j \mid \boldsymbol{w, \theta}) = p(y_j \mid \theta_c) \mathds{I} (c \in A(u_j \mid \boldsymbol{w})),
\end{align*}
where $A(u_j \mid \boldsymbol{w}) = \{ c : w_c > u_j\}$. 
The joint likelihood is given by
\begin{align*}
f(\boldsymbol{y, u, z}) = \prod_{j} p(y_j \mid \theta_{z_j}) \mathds{I}(u_j < w_{z_j}).
\end{align*}

The full conditional distributions are described as follows:
\begin{itemize}
\item[-] (Sampling of $\theta^\star_c$)
For the parameter $\theta^\star_c$ of the $c$th cluster, we obtain 
\begin{align*}
p (\theta^\star_c \mid \ldots) \propto g_0 (\theta^\star_c) \prod_{z_j = c} p(y_j \mid \theta_c)
\end{align*}
where $g_0 (\theta^\star_c)$ denotes the base measure, as in \citet{walkerSamplingDirichletMixture2007}.
If the $c$ th cluster is empty, we also obtain
\begin{align*}
p (\theta^\star_c \mid \ldots) \propto g_0 (\theta_c) .
\end{align*}

\item[-] (Sampling of $u_j$)
For $j \left(= \text{H}_1, \ldots, \text{H}_K, \text{CC}\right)$, the full conditional distribution of $u_j$ is given by $\text{Unif} (0, w_{z_j})$.

\item[-] (Sampling of $w$)
The variable $w$ is constructed from $v$ and illustrates the sampling procedure for 
$v$.
For the parameter $v_c (c = 1,\ldots, r)$, we obtain
\begin{align*}
p(v_c \mid \ldots) &\propto p(v_c) \prod_{j} \mathds{I} (u_j < w_{z_j}) \\
 & =  p(v_c) \prod_{j} \mathds{I} \left( v_{z_j} \prod_{\ell < z_j} (1-v_\ell) > u_j \right).
\end{align*}
Let $c^\ast$ denote the maximum number of cluster numbership of $z_j$.
Consequently, for $c > c^\ast$, the full conditional distribution of $v_c$ is given by $\text{Beta}(1,M)$.

\item[-] (Sampling of $z_j$)
For $j \left(= \text{H}_1, \ldots, \text{H}_K, \text{CC}\right)$, the full conditional distribution of the cluster membership variable $z_j$ is given by
\begin{align*}
\text{Pr}(z_j = c \mid \ldots) \propto \mathds{I} (c \in A_w (u_j)) p(y_j \mid \theta_{z_j}),
\end{align*}
where the range of $c$ is defined as the minimum value of $c^\ast$ that satisfies the following condition:
\begin{align*}
\sum_{c=1}^{c^\ast} w_c > 1 - u^\ast,
\end{align*}
with $u^\ast = \min{u_{\text{H}_1}, \ldots, u_{\text{H}_K}, u_{\text{CC}}}$.

\item[-] (Sampling of $M$)
The concentration parameter $M$ is updated as in \citet{escobarBayesianDensityEstimation1995} assuming a gamma prior $\text{Ga}(a, b)$.

\end{itemize}

\section{Step-by-step Sampling Procedures for dependent Dirichlet process mixture method}\label{MCMC_DDPM}
As in Section \ref{MCMC_DPM}, we show the posterior sampling procedure when the outcome data $y$ are aggregated study-level data, for simplicity.
The joint likelihood is given by
\begin{align*}
f(\boldsymbol{y, u, z} \mid \boldsymbol{w}^{(\text{H})}, \boldsymbol{w}^{(\text{CC})}, \boldsymbol{\theta}) &= \prod_{j \in (\text{H}_1, \ldots, \text{H}_K)} \mathds{I}(u_j < w_{z_j}^{(\text{H})}) p(y_j \mid \theta_{z_j}) \\
&\times \mathds{I}(u_\text{CC} < w_{z_\text{CC}}^{(\text{CC})}) p(y_\text{CC} \mid \theta_{z_\text{CC}})
\end{align*}
Next, we present the sampling procedure for the newly introduced variable $\phi$ in the DDPM method and 
$w_{z_j}^{(\text{H})}, w_{z_j}^{(\text{CC})}$, which follows a different sampling procedure from that in the DPM method.

The full conditional distributions are described as follows:
\begin{itemize}
\item[-] (Sampling of $w$)
To update the weights, we need to update the stick-breaking components; to this end, denote by $p_c^{(\text{CC})}$ the transition density $\mathds{P} (V_c^{(\text{CC})} \in A \mid V_c^{(\text{H})} )$ correspounding to the $c$th processs.
Hence, it is seen that
\begin{align*}
p(v_c^{(\cdot)} \mid \ldots) \propto \left\{ p_c^{(\text{CC})} \text{Beta}(v_c^{(\text{H})} ; 1, M) \mathds{I}(j \in \text{H}) + p_c^{(\text{CC})} \mathds{I} (j \in \text{CC}) \right\} \times v_c^{(\cdot)n_c^{(\cdot)}} (1 - v_c^{(\cdot)})^{m_c^{(\cdot)}}
\end{align*}
where $\text{Beta} (\cdot ; a, b)$ denotes the density of a Beta distirbution and 
\begin{align*}
n_c^{(\cdot)} := \sum_{j} \mathds{I}(z_j = c), \ \ m_c^{(\cdot)} := \sum_{j} \mathds{I}(z_j > c).
\end{align*}
For the weights for historical controls, we obtain
\begin{align*}
p(v_c^{(\text{H})}) &= M \times \frac{\phi M (1 - v_j^{(\text{CC})})^{M-1}}{B(1+n_j^{(\text{H})}, M + m_j^{(\text{H})})} \text{Beta} (v_j^{(\text{H})}; 1 + n_j^{(\text{H})}, M + m_j^{(\text{H})}) \\
&+ M \times (1 - \phi) (1 - v_j^{(\text{CC})})^{M-1} (v_j^{(\text{CC})})^{n_j^{(\text{H})}} (1 - v_j^{(\text{CC})})^{m_j^{(\text{H})}} \mathds{I} (v_j^{(\text{H})} = v_j^{(\text{CC})}),
\end{align*}
with $B(a, b) = \Gamma(a+b)/(\Gamma(a) \Gamma(b))$.
For the weights for current control, we also obtain
\begin{align*}
p(v_c^{(\text{CC})}) &= \frac{\phi M (1 - 0)^{M-1}}{B(1 + n_j^{(\text{CC})}, M + m_j^{(\text{CC})})} \times \text{Beta}(v_j^{(\text{CC})}; 1 + n_j^{(\text{CC})}, M + m_j^{(\text{CC})}) \\
&+ (1-\phi) (1 - 0)^{M-1} (v_j^{(\text{H})})^{n_j^{(\text{CC})}} (1 - v_j^{(\text{H})})^{m_j^{(\text{CC})}} \mathds{I}(v_j^{(\text{CC})} = v_j^{(\text{H})}).
\end{align*}

\item[-] (Sampling of $\phi$)
The posterior distribution for $\phi$ is not available in closed form, thus a Metropolis--Hasting step is needed \citep{gutierrezTimeDependentBayesian2016}.
As procedure in \citet{gutierrezTimeDependentBayesian2016}, we use a truncated normal distribution as a proposal for $\phi$, that is , at iteration $s$, $\phi^\ast \sim \text{N} (\phi^\ast \mid \phi^{s-1}, c) \mathds{I}_{[0, 1]}$.
Then set:
\begin{align*}
\phi^s = 
\begin{cases}
\phi^\ast & \text{with probability} \min(r, 1) \\
\phi^{s-1} & \text{otherwise}
\end{cases}
\end{align*}
and
\begin{align*}
r = \frac{p(\phi^\ast \mid \boldsymbol{y}) / \text{N}(\phi^\ast \mid \phi^{s-1}, c) \mathds{I}_{[0,1]} }{p(\phi^{s-1} \mid \boldsymbol{y}) / \text{N}(\phi^{s-1} \mid \phi^\ast, c) \mathds{I}_{[0,1]} },
\end{align*}
with $p(\phi^\diamond \mid \boldsymbol{y})$ is given by
\begin{align*}
p(\phi^\diamond \mid \boldsymbol{y}) \propto \phi^{\diamond (\gamma - 1)} (1 - \phi^{\diamond})^{(\nu - 1)} \prod_{j} \prod_{c} w_c^\diamond p(y_j \mid \theta_c)
\end{align*}
and $w_j^\diamond$ are the weights sampled using $\phi^\diamond$.

\end{itemize}

\newpage

\section{Proof of theorem}\label{proof}
\subsection{Proof of theorem 1}
As demonstrated in Theorem 1 in \citet{ascolaniClusteringConsistencyDirichlet2023}, if the prior distribution for the concentration parameter $M$ satisfies assumptions 1--3 in \citet{ascolaniClusteringConsistencyDirichlet2023}, then as the number of historical controls increases, the posterior of the number of clusters is consistent with the true number of clusters.

\subsection{Proof of theorem 2}
As stated in Theorem 6.2 of \cite{ghosalFundamentalsNonparametricBayesian2017}, if the probability measure under the prior distribution is positive, the posterior distribution for each study's parameter $\theta_j (j = \text{H}_1, \ldots, \text{H}_K, \text{CC})$ converges to the true parameter as $n_j \rightarrow \infty$.
In other words, the posterior distribution converges to the Dirac measure concentrated on the true parameter.

Meanwhile, the conditional posterior distribution of DPM is expressed as:
\begin{align*}
p(\theta_j \mid \boldsymbol{\theta}_{-j}) \propto \sum_{j' = 1}^{r^{-}} n_{j'}^{-} \delta_{\theta_{j'}^{\star -}} (\theta_j) + M G_0, (\theta_j)
\end{align*}
where $\boldsymbol{\theta}_{-j}$ denotes $\boldsymbol{\theta}$ without the $j$th element $\theta_j$, $r^{-}$ is the number of unique values in $\boldsymbol{\theta}_{-j}$ and $\theta_{j'}^{\star -}$ is the $j'$th unique element.
Based on this, the posterior distribution converges to the Dirac measure concentrated on the true parameter as $n_j \rightarrow \infty$.
Consequently, the posterior distributions of $\theta_{\text{CC}}$ and $\theta_{\text{H}_k}$ converge to the same true value if the current control and the $k$th historical control are congruent, and to distinct values if the current control and the $k$th historical control are incongruent.

\newpage

\section{Results of simulation study for summary statistics}
\begin{table}[htbp]
	\caption{Bias of the posterior mean of the posterior distribution of treatment effect $\pi_\text{CT} - \pi_\text{CC} (\%)$ in the simulation study for summary statistics. In DPM and DDPM methods, the numbers in parentheses indicate the shape and scale parameters of the gamma distribution for the concentration parameter $M$.}
	\centering
	\begin{tabular}{lrrrrr}
		\toprule
		Method  & Sce 1 & Sce 2 & Sce 3 & Sce 4 & Sce 5 \\
		\midrule
		CD         & $-$0.78 & $-$1.58 & $-$0.85 & $-$1.51 & $-$0.17 \\
		PD         & $-$0.64 & $-$1.63 & 6.50  & 13.71 & 28.14 \\
		MAP        & $-$0.67 & $-$1.56 & 1.81  & 2.87  & 17.28 \\
		DPM-MAP    & $-$0.70 & $-$1.56 & 0.91  & 2.98  & 9.65  \\
		SAM-MAP    & $-$1.61 & $-$2.36 & $-$0.59 & $-$1.29 & $-$0.15 \\
		DPM (1, 1)   & $-$0.65 & $-$1.51 & 0.74  & 1.42  & 12.07 \\
		DPM (1, 5)   & $-$0.66 & $-$1.51 & 0.61  & 1.23  & 10.71 \\
		DPM (1, 10)  & $-$0.66 & $-$1.51 & 0.59  & 1.19  & 10.47 \\
		DDPM (1, 1)  & $-$0.66 & $-$1.52 & 0.89  & 1.25  & 9.47  \\
		DDPM (1, 5)  & $-$0.67 & $-$1.53 & 0.78  & 1.11  & 8.63  \\
		DDPM (1, 10) & $-$0.67 & $-$1.53 & 0.76  & 1.08  & 8.49 \\
		\bottomrule
	\end{tabular}	
	\begin{tablenotes}
		\item Abbreviations: CD, current data analysis; PD, pooled data analysis; MAP, meta-analytic predictive prior method; DPM-MAP, Dirichlet process mixture-meta-analytic predictive prior method; SAM-MAP, self-adapting mixture-meta-analytic predictive prior method; DPM, Dirichlet process mixture method; DDPM, dependent Dirichlet process mixture method.
		\item Note: The five scenarios are defined as follows: Sce 1, no between-trial heterogeneity; Sce 2, large between-trial heterogeneity; Sce 3, two heterogeneous historical controls exist; Sce 4, four heterogeneous historical controls exist; Sce 5, eight heterogeneous historical controls exist.
	\end{tablenotes}
\end{table}
\begin{table}[htbp]
	\caption{Root mean square error of the posterior mean of the posterior distribution of treatment effect $\pi_\text{CT} - \pi_\text{CC} (\%)$ in the simulation study for summary statistics. In DPM and DDPM methods, the numbers in parentheses indicate the shape and scale parameters of the gamma distribution for the concentration parameter $M$.}
	\centering
	\begin{tabular}{lrrrrr}
		\toprule
		Method  & Sce 1 & Sce 2 & Sce 3 & Sce 4 & Sce 5 \\
		\midrule
		CD           & 12.60 & 12.49 & 12.67 & 12.28 & 12.78 \\
		PD           & 7.13  & 11.93 & 9.60  & 15.41 & 28.98 \\
		MAP          & 7.54  & 10.18 & 10.48 & 11.03 & 20.24 \\
		DPM-MAP      & 8.21  & 10.32 & 9.83  & 11.07 & 17.09 \\
		SAM-MAP      & 10.61 & 11.78 & 12.34 & 12.23 & 12.80 \\
		DPM (1, 1)   & 7.25  & 10.46 & 8.82  & 9.94  & 21.24 \\
		DPM (1, 5)   & 7.32  & 10.49 & 8.97  & 10.07 & 20.37 \\
		DPM (1, 10)  & 7.34  & 10.50 & 9.01  & 10.10 & 20.22 \\
		DDPM (1, 1)  & 7.36  & 10.55 & 9.14  & 9.96  & 19.75 \\
		DDPM (1, 5)  & 7.44  & 10.57 & 9.25  & 10.06 & 19.18 \\
		DDPM (1, 10) & 7.45  & 10.58 & 9.27  & 10.08 & 19.08 \\
		\bottomrule
	\end{tabular}	
	\begin{tablenotes}
		\item Abbreviations: CD, current data analysis; PD, pooled data analysis; MAP, meta-analytic predictive prior method; DPM-MAP, Dirichlet process mixture-meta-analytic predictive prior method; SAM-MAP, self-adapting mixture-meta-analytic predictive prior method; DPM, Dirichlet process mixture method; DDPM, dependent Dirichlet process mixture method.
		\item Note: The five scenarios are defined as follows: Sce 1, no between-trial heterogeneity; Sce 2, large between-trial heterogeneity; Sce 3, two heterogeneous historical controls exist; Sce 4, four heterogeneous historical controls exist; Sce 5, eight heterogeneous historical controls exist.
	\end{tablenotes}
\end{table}
\begin{table}[htbp]
	\caption{Mean posterior standard deviation of the posterior mean of the posterior distribution of treatment effect $\pi_\text{CT} - \pi_\text{CC} (\%)$ in the simulation study for summary statistics. In DPM and DDPM methods, the numbers in parentheses indicate the shape and scale parameters of the gamma distribution for the concentration parameter $M$.}
	\centering
	\begin{tabular}{lrrrrr}
		\toprule
		Method  & Sce 1 & Sce 2 & Sce 3 & Sce 4 & Sce 5 \\
		\midrule
		CD           & 12.38 & 12.12 & 12.38 & 12.39 & 12.37 \\
		PD           & 7.03  & 6.94  & 7.03  & 7.01  & 6.92  \\
		MAP          & 8.48  & 10.64 & 11.19 & 11.55 & 10.49 \\
		DPM-MAP      & 9.45  & 10.83 & 10.81 & 11.53 & 12.58 \\
		SAM-MAP      & 11.26 & 11.75 & 12.18 & 12.36 & 12.37 \\
		DPM (1, 1)   & 7.63  & 10.15 & 9.39  & 10.03 & 13.00 \\
		DPM (1, 5)   & 7.84  & 10.35 & 9.63  & 10.27 & 13.18 \\
		DPM (1, 10)  & 7.88  & 10.39 & 9.67  & 10.32 & 13.21 \\
		DDPM (1, 1)  & 7.85  & 10.18 & 9.49  & 9.89  & 13.34 \\
		DDPM (1, 5)  & 8.03  & 10.33 & 9.68  & 10.10 & 13.39 \\
		DDPM (1, 10) & 8.06  & 10.36 & 9.72  & 10.14 & 13.39
		\\
		\bottomrule
	\end{tabular}	
	\begin{tablenotes}
		\item Abbreviations: CD, current data analysis; PD, pooled data analysis; MAP, meta-analytic predictive prior method; DPM-MAP, Dirichlet process mixture-meta-analytic predictive prior method; SAM-MAP, self-adapting mixture-meta-analytic predictive prior method; DPM, Dirichlet process mixture method; DDPM, dependent Dirichlet process mixture method.
		\item Note: The five scenarios are defined as follows: Sce 1, no between-trial heterogeneity; Sce 2, large between-trial heterogeneity; Sce 3, two heterogeneous historical controls exist; Sce 4, four heterogeneous historical controls exist; Sce 5, eight heterogeneous historical controls exist.
	\end{tablenotes}
\end{table}
\begin{table}[htbp]
	\caption{Coverage probability (\%) of the posterior mean of the posterior distribution of treatment effect in the simulation study for summary statistics. In DPM and DDPM methods, the numbers in parentheses indicate the shape and scale parameters of the gamma distribution for the concentration parameter $M$.}
	\centering
	\begin{tabular}{lrrrrr}
		\toprule
		Method  & Sce 1 & Sce 2 & Sce 3 & Sce 4 & Sce 5 \\
		\midrule
		CD           & 95.11 & 94.31 & 94.80 & 95.43 & 94.72 \\
		PD           & 94.70 & 76.25 & 84.79 & 54.52 & 3.61  \\
		MAP          & 97.27 & 95.90 & 96.52 & 95.82 & 63.66 \\
		DPM-MAP      & 97.55 & 96.02 & 96.96 & 95.71 & 83.96 \\
		SAM-MAP      & 96.60 & 94.71 & 94.75 & 95.30 & 94.71 \\
		DPM (1, 1)   & 95.82 & 93.92 & 96.47 & 95.71 & 72.16 \\
		DPM (1, 5)   & 96.11 & 94.47 & 96.59 & 96.02 & 76.15 \\
		DPM (1, 10)  & 96.11 & 94.52 & 96.60 & 96.08 & 76.69 \\
		DDPM (1, 1)  & 95.91 & 93.91 & 96.24 & 95.56 & 78.81 \\
		DDPM (1, 5)  & 96.14 & 94.31 & 96.32 & 95.82 & 80.54 \\
		DDPM (1, 10) & 96.18 & 94.34 & 96.36 & 95.86 & 80.89
		\\
		\bottomrule
	\end{tabular}	
	\begin{tablenotes}
		\item Abbreviations: CD, current data analysis; PD, pooled data analysis; MAP, meta-analytic predictive prior method; DPM-MAP, Dirichlet process mixture-meta-analytic predictive prior method; SAM-MAP, self-adapting mixture-meta-analytic predictive prior method; DPM, Dirichlet process mixture method; DDPM, dependent Dirichlet process mixture method.
		\item Note: The five scenarios are defined as follows: Sce 1, no between-trial heterogeneity; Sce 2, large between-trial heterogeneity; Sce 3, two heterogeneous historical controls exist; Sce 4, four heterogeneous historical controls exist; Sce 5, eight heterogeneous historical controls exist.
	\end{tablenotes}
\end{table}
\begin{table}[htbp]
	\caption{Type I error rate (\%) of the posterior mean of the posterior distribution of treatment effect in the simulation study for summary statistics. In DPM and DDPM methods, the numbers in parentheses indicate the shape and scale parameters of the gamma distribution for the concentration parameter $M$.}
	\centering
	\begin{tabular}{lrrrrr}
		\toprule
		Method  & Sce 1 & Sce 2 & Sce 3 & Sce 4 & Sce 5 \\
		\midrule
		CD           & 2.33 & 2.60  & 2.61  & 2.53  & 2.84  \\
		PD           & 2.65 & 13.87 & 14.81 & 44.46 & 96.66 \\
		MAP          & 1.65 & 2.80  & 3.17  & 4.17  & 36.93 \\
		DPM-MAP      & 1.49 & 2.59  & 2.45  & 4.34  & 15.12 \\
		SAM-MAP      & 1.31 & 2.07  & 2.53  & 2.53  & 2.84  \\
		DPM (1, 1)   & 2.24 & 3.90  & 2.60  & 3.45  & 25.81 \\
		DPM (1, 5)   & 2.16 & 3.66  & 2.44  & 3.27  & 22.18 \\
		DPM (1, 10)  & 2.12 & 3.59  & 2.43  & 3.28  & 21.53 \\
		DDPM (1, 1)  & 2.14 & 3.83  & 2.85  & 3.55  & 19.21 \\
		DDPM (1, 5)  & 2.07 & 3.56  & 2.70  & 3.40  & 17.03 \\
		DDPM (1, 10) & 2.05 & 3.59  & 2.71  & 3.38  & 16.76
		\\
		\bottomrule
	\end{tabular}	
	\begin{tablenotes}
		\item Abbreviations: CD, current data analysis; PD, pooled data analysis; MAP, meta-analytic predictive prior method; DPM-MAP, Dirichlet process mixture-meta-analytic predictive prior method; SAM-MAP, self-adapting mixture-meta-analytic predictive prior method; DPM, Dirichlet process mixture method; DDPM, dependent Dirichlet process mixture method.
		\item Note: The five scenarios are defined as follows: Sce 1, no between-trial heterogeneity; Sce 2, large between-trial heterogeneity; Sce 3, two heterogeneous historical controls exist; Sce 4, four heterogeneous historical controls exist; Sce 5, eight heterogeneous historical controls exist.
	\end{tablenotes}
\end{table}
\begin{table}[htbp]
	\caption{Power (\%) of the posterior mean of the posterior distribution of treatment effect in the simulation study for summary statistics. In DPM and DDPM methods, the numbers in parentheses indicate the shape and scale parameters of the gamma distribution for the concentration parameter $M$.}
	\centering
	\begin{tabular}{lrrrrr}
		\toprule
		Method  & Sce 1 & Sce 2 & Sce 3 & Sce 4 & Sce 5 \\
		\midrule
		CD           & 47.19 & 46.50 & 46.86 & 44.30 & 49.06  \\
		PD           & 87.90 & 74.93 & 98.14 & 99.87 & 100.00 \\
		MAP          & 78.38 & 56.99 & 63.93 & 64.16 & 95.17  \\
		DPM-MAP      & 70.34 & 55.63 & 64.88 & 64.23 & 69.50  \\
		SAM-MAP      & 55.05 & 47.23 & 49.56 & 44.67 & 49.06  \\
		DPM (1, 1)   & 83.95 & 60.19 & 78.70 & 75.62 & 63.78  \\
		DPM (1, 5)   & 82.42 & 58.92 & 76.18 & 72.19 & 61.58  \\
		DPM (1, 10)  & 82.34 & 58.62 & 75.64 & 71.58 & 61.20  \\
		DDPM (1, 1)  & 82.40 & 59.89 & 78.16 & 75.49 & 59.41  \\
		DDPM (1, 5)  & 80.83 & 58.91 & 76.29 & 72.81 & 58.32  \\
		DDPM (1, 10) & 80.68 & 58.77 & 75.70 & 72.43 & 58.17
		\\
		\bottomrule
	\end{tabular}	
	\begin{tablenotes}
		\item Abbreviations: CD, current data analysis; PD, pooled data analysis; MAP, meta-analytic predictive prior method; DPM-MAP, Dirichlet process mixture-meta-analytic predictive prior method; SAM-MAP, self-adapting mixture-meta-analytic predictive prior method; DPM, Dirichlet process mixture method; DDPM, dependent Dirichlet process mixture method.
		\item Note: The five scenarios are defined as follows: Sce 1, no between-trial heterogeneity; Sce 2, large between-trial heterogeneity; Sce 3, two heterogeneous historical controls exist; Sce 4, four heterogeneous historical controls exist; Sce 5, eight heterogeneous historical controls exist.
	\end{tablenotes}
\end{table}
\begin{table}[htbp]
	\caption{mean effective historical sample size of the posterior distribution of treatment effect in the simulation study for summary statistics. In DPM and DDPM methods, the numbers in parentheses indicate the shape and scale parameters of the gamma distribution for the concentration parameter $M$.}
	\centering
	\begin{tabular}{lrrrrr}
		\toprule
		Method  & Sce 1 & Sce 2 & Sce 3 & Sce 4 & Sce 5 \\
		\midrule
		MAP          & 123.2 & 21.8  & 10.4  & 7.1   & 65.0  \\
		DPM-MAP      & 62.9  & 17.9  & 16.5  & 7.5   & 20.1  \\
		SAM-MAP      & 36.6  & 6.2   & 2.5   & 1.5   & 1.3   \\
		DPM (1, 1)   & 373.9 & 115.8 & 197.7 & 129.2 & 192.7 \\
		DPM (1, 5)   & 342.8 & 95.2  & 167.2 & 106.7 & 159.5 \\
		DPM (1, 10)  & 336.9 & 91.6  & 161.3 & 102.5 & 154.0 \\
		DDPM (1, 1)  & 394.5 & 130.0 & 210.2 & 143.5 & 174.2 \\
		DDPM (1, 5)  & 372.8 & 113.7 & 187.7 & 125.1 & 152.0 \\
		DDPM (1, 10) & 368.7 & 110.7 & 183.4 & 121.3 & 148.4
		\\
		\bottomrule
	\end{tabular}	
	\begin{tablenotes}
		\item Abbreviations: MAP, meta-analytic predictive prior method; DPM-MAP, Dirichlet process mixture-meta-analytic predictive prior method; SAM-MAP, self-adapting mixture-meta-analytic predictive prior method; DPM, Dirichlet process mixture method; DDPM, dependent Dirichlet process mixture method.
		\item Note: The five scenarios are defined as follows: Sce 1, no between-trial heterogeneity; Sce 2, large between-trial heterogeneity; Sce 3, two heterogeneous historical controls exist; Sce 4, four heterogeneous historical controls exist; Sce 5, eight heterogeneous historical controls exist.
	\end{tablenotes}
\end{table}
\begin{table}[htbp]
	\caption{Standard deviation of effective historical sample size of the posterior distribution of treatment effect in the simulation study for summary statistics. In DPM and DDPM methods, the numbers in parentheses indicate the shape and scale parameters of the gamma distribution for the concentration parameter $M$.}
	\centering
	\begin{tabular}{lrrrrr}
		\toprule
		Method  & Sce 1 & Sce 2 & Sce 3 & Sce 4 & Sce 5 \\
		\midrule
		MAP          & 43.7 & 25.3  & 5.1  & 3.3  & 47.4  \\
		DPM-MAP      & 21.3 & 13.9  & 6.4  & 3.7  & 21.6  \\
		SAM-MAP      & 22.2 & 9.2   & 1.3  & 0.4  & 0.1   \\
		DPM (1, 1)   & 71.7 & 94.3  & 69.5 & 44.6 & 149.1 \\
		DPM (1, 5)   & 79.4 & 83.6  & 65.9 & 42.1 & 137.5 \\
		DPM (1, 10)  & 80.7 & 81.7  & 64.9 & 41.3 & 135.3 \\
		DDPM (1, 1)  & 62.8 & 103.6 & 70.3 & 44.0 & 150.6 \\
		DDPM (1, 5)  & 69.4 & 95.8  & 68.1 & 43.0 & 140.9 \\
		DDPM (1, 10) & 70.5 & 94.4  & 67.6 & 42.7 & 139.2
		\\
		\bottomrule
	\end{tabular}	
	\begin{tablenotes}
		\item Abbreviations: MAP, meta-analytic predictive prior method; DPM-MAP, Dirichlet process mixture-meta-analytic predictive prior method; SAM-MAP, self-adapting mixture-meta-analytic predictive prior method; DPM, Dirichlet process mixture method; DDPM, dependent Dirichlet process mixture method.
		\item Note: The five scenarios are defined as follows: Sce 1, no between-trial heterogeneity; Sce 2, large between-trial heterogeneity; Sce 3, two heterogeneous historical controls exist; Sce 4, four heterogeneous historical controls exist; Sce 5, eight heterogeneous historical controls exist.
	\end{tablenotes}
\end{table}

\newpage
\section{Results of simulation study for individual participant data}
\begin{table}[htbp]
	\caption{Root mean square error of the posterior mean of the posterior distribution of treatment effect $\beta_\text{TRT}$ in the simulation study for individual participant data.}
	\centering
	\begin{tabular}{lrrrr}
		\toprule
		Method  & Sce 1 & Sce 2 & Sce 3 & Sce 4 \\
		\midrule
		CD   & 1.46 & 1.48 & 1.46 & 1.49 \\
		PD   & 1.11 & 3.06 & 1.53 & 2.43 \\
		EX   & 1.19 & 1.45 & 1.39 & 1.46 \\
		DPM  & 1.11 & 2.11 & 1.15 & 1.15 \\
		DDPM & 1.11 & 1.92 & 1.14 & 1.14 \\
		\bottomrule
	\end{tabular}	
	\begin{tablenotes}
		\item Abbreviations: CD, current data analysis; PD, pooled data analysis; EX, exchangeable method that assumes partial exchangeability for the intercept; DPM, Dirichlet process mixture method; DDPM, dependent Dirichlet process mixture method.
		\item Note: The four scenarios are defined as follows: Sce 1, no between-trial heterogeneity; Sce 2, large between-trial heterogeneity; Sce 3, one heterogeneous historical control exists; Sce 4, two heterogeneous historical controls exist.		
	\end{tablenotes}
\end{table}

\begin{table}[htbp]
	\caption{Mean posterior standard deviation of the posterior mean of the posterior distribution of treatment effect $\beta_\text{TRT}$ in the simulation study for individual participant data.}
	\centering
	\begin{tabular}{lrrrr}
		\toprule
		Method  & Sce 1 & Sce 2 & Sce 3 & Sce 4 \\
		\midrule
		CD   & 1.50 & 1.50 & 1.49 & 1.49 \\
		PD   & 1.10 & 1.16 & 1.14 & 1.16 \\
		EX   & 1.28 & 1.44 & 1.44 & 1.45 \\
		DPM  & 1.10 & 1.17 & 1.12 & 1.13 \\
		DDPM & 1.10 & 1.18 & 1.12 & 1.13 \\
		\bottomrule
	\end{tabular}	
	\begin{tablenotes}
		\item Abbreviations: CD, current data analysis; PD, pooled data analysis; EX, exchangeable method that assumes partial exchangeability for the intercept; DPM, Dirichlet process mixture method; DDPM, dependent Dirichlet process mixture method.
		\item Note: The four scenarios are defined as follows: Sce 1, no between-trial heterogeneity; Sce 2, large between-trial heterogeneity; Sce 3, one heterogeneous historical control exists; Sce 4, two heterogeneous historical controls exist.		
	\end{tablenotes}
\end{table}

\begin{table}[htbp]
	\caption{Coverage probability (\%) of the posterior mean of the posterior distribution of treatment effect $\beta_\text{TRT}$ in the simulation study for individual participant data.}
	\centering
	\begin{tabular}{lrrrr}
		\toprule
		Method  & Sce 1 & Sce 2 & Sce 3 & Sce 4 \\
		\midrule
		CD   & 95.8 & 95.3 & 94.9 & 94.8 \\
		PD   & 94.9 & 52.8 & 85.7 & 53.2 \\
		EX   & 96.7 & 94.9 & 95.9 & 94.9 \\
		DPM  & 95.2 & 71.6 & 94.1 & 94.1 \\
		DDPM & 95.2 & 76.7 & 94.7 & 94.5 \\
		\bottomrule
	\end{tabular}	
	\begin{tablenotes}
		\item Abbreviations: CD, current data analysis; PD, pooled data analysis; EX, exchangeable method that assumes partial exchangeability for the intercept; DPM, Dirichlet process mixture method; DDPM, dependent Dirichlet process mixture method.
		\item Note: The four scenarios are defined as follows: Sce 1, no between-trial heterogeneity; Sce 2, large between-trial heterogeneity; Sce 3, one heterogeneous historical control exists; Sce 4, two heterogeneous historical controls exist.		
	\end{tablenotes}
\end{table}